\documentclass[conference]{IEEEtran}

\usepackage{cite}
\usepackage{amsmath,amssymb,amsfonts}
\usepackage{algorithmic}
\usepackage{graphicx}
\usepackage{textcomp}
\usepackage{xcolor}
\usepackage{algorithm2e}

\newcommand{\ve}[1]{\mathbf{#1}} % or \bm
\usepackage[nolist]{acronym}
\usepackage[nolist]{acronym}

\newacro{GNSS}{Global Navigation Satellite System}
\newacro{LSTM NN}{long-short term memory neural network}
\newacro{ML}{machine learning}
\newacro{WLS}{weighted least squares}
\newacro{FDE}{fault detection and exclusion}
\newacro{RNN}{recurrent neural network}
\newacro{MSE}{mean squared error}
\newacro{CRLB}{Cramer Rao lower bound}
\newacro{CDF}{cumulative density function}
\newacro{SV}{satellite vehicle}

\def\BibTeX{{\rm B\kern-.05em{\sc i\kern-.025em b}\kern-.08em
    T\kern-.1667em\lower.7ex\hbox{E}\kern-.125emX}}
\begin{document}

\title{A Novel Satellite Selection Algorithm Using LSTM Neural Networks For Single-epoch Localization}

\author{\IEEEauthorblockN{Ibrahim Sbeity}
\IEEEauthorblockA{\textit{CEA-Leti} \\
\textit{Universit\'e Grenoble Alpes}\\
F-38000 Grenoble, France \\
\textit{ETIS UMR 8051} \\
\textit{CY Cergy Paris Universit\'e, ENSEA, CNRS}\\
F-95000 Cergy, France \\
Ibrahim.Sbeity@cea.fr}
\and
\IEEEauthorblockN{Christophe Villien}
\IEEEauthorblockA{\textit{CEA-Leti} \\
\textit{Universit\'e Grenoble Alpes}\\
F-38000 Grenoble, France \\
Christophe.Villien@cea.fr}
\and
\IEEEauthorblockN{Marwa Chafii}
\IEEEauthorblockA{\textit{Engineering Division} \\
\textit{New York University (NYU) Abu Dhabi}\\
Abu Dhabi, UAE \\
\textit{NYU WIRELESS} \\
\textit{NYU Tandon School of Engineering}\\
Brooklyn, NY 11201, USA \\
email address or ORCID}
\and
\IEEEauthorblockN{Christophe Combettes}
\IEEEauthorblockA{\textit{CEA-Leti} \\
\textit{Universit\'e Grenoble Alpes}\\
F-38000 Grenoble, France \\
Christophe.Combettes@cea.fr}
\and
\IEEEauthorblockN{Benoît Denis}
\IEEEauthorblockA{\textit{CEA-Leti} \\
\textit{Universit\'e Grenoble Alpes}\\
F-38000 Grenoble, France \\
Benoit.Denis@cea.fr}
\and
\IEEEauthorblockN{E. Veronica Belmega}
\IEEEauthorblockA{\textit{dept. name of organization (of Aff.)} \\
\textit{Univ. Gustave Eiffel, CNRS, LIGM}\\
F-77454 Marne-la-Valle\'e, France \\
\textit{ETIS UMR 8051} \\
\textit{CY Cergy Paris Universit\'e, ENSEA, CNRS}\\
F-95000 Cergy, France \\
email address or ORCID}
}

\author{\IEEEauthorblockN{Ibrahim Sbeity\IEEEauthorrefmark{1}\IEEEauthorrefmark{2},
Christophe Villien\IEEEauthorrefmark{1}, Christophe Combettes\IEEEauthorrefmark{1},\\ Benoît Denis\IEEEauthorrefmark{1}, E. Veronica Belmega\IEEEauthorrefmark{3}\IEEEauthorrefmark{2}, and Marwa Chafii \IEEEauthorrefmark{4}\IEEEauthorrefmark{5}}\\
\IEEEauthorblockA{
\IEEEauthorrefmark{1}
CEA-Leti, Universit\'e Grenoble Alpes, 
F-38000 Grenoble, France \\
\IEEEauthorrefmark{2} ETIS UMR 8051, CY Cergy Paris Universit\'e, ENSEA, CNRS, F-95000, Cergy, France\\
\IEEEauthorrefmark{3} Univ. Gustave Eiffel, CNRS, LIGM, F-77454,  Marne-la-Vallée, France\\
\IEEEauthorrefmark{4}
New York University (NYU) Abu Dhabi,
Abu Dhabi, UAE \\
\IEEEauthorrefmark{5}
NYU WIRELESS, NYU Tandon School of Engineering,
Brooklyn, NY 11201, USA \\ \\
Emails: ibrahim.sbeity@cea.fr, christophe.villien@cea.fr}
%\thanks{This work has been supported by TBD.}
}

\maketitle

\begin{abstract}
This work presents a new approach for detection and exclusion (or de-weighting) of pseudo-range measurements from the \ac{GNSS} in order to improve the accuracy of single-epoch positioning, which is an essential prerequisite for maintaining good navigation performance in challenging operating contexts (e.g., under Non-Line of Sight and/or multipath propagation). Beyond the usual preliminary hard decision stage, which can mainly reject obvious outliers, our approach exploits machine learning to optimize the relative contributions from all available satellites feeding the positioning solver. For this, we construct a customized matrix of pseudo-range residuals that is used as an input to the proposed \ac{LSTM NN} architecture. The latter is trained to predict several quality indicators that roughly approximate the standard deviations of pseudo-range errors, which are further integrated in the calculation of weights. Our numerical evaluations on both synthetic and real data show that the proposed solution is able to outperform conventional weighting and signal selection strategies from the state-of-the-art, while fairly approaching optimal positioning accuracy.    
\end{abstract}


\section{Introduction}
Accurate and resilient positioning services based on the \acf{GNSS} have become essential into a variety of outdoor applications, such as autonomous vehicles and unmanned aerial vehicles, blueforce and first responders tracking, seamless end-to-end logistics and supply chains optimization, large-scale crowd-sensing, etc. 

By itself, single-epoch \ac{GNSS} localization is a key enabler that can provide tracking filters with input observations so as to refine further the mobile position and exploit its dynamics through hybrid data fusion (e.g., combining \ac{GNSS} with other modalities from inertial sensors, odometers, etc.). Beyond, single-epoch localization is also and foremost typically used to initialize the navigation processor in charge of tracking the \ac{GNSS} filtered solution \cite{grewal2007global}. 

The signals received from satellites can be severely affected by Non-Line of Sight (NLOS) and multipath (MP) propagation in harsh environments, such as urban canyons, hence leading to strongly biased pseudo-range measurements. 
In this context, selecting the most reliable and/or most informative measurements (while discarding the most harmful ones) is of primary importance to preserve the accuracy of this preliminary single-epoch positioning stage and, hence, of the overall navigation system. This down-selection step is even more critical and challenging as several tens of measurements are typically made available within modern \ac{GNSS} receivers at each time epoch (i.e., while considering multiple satellites from different constellations), making exhaustive search computationally prohibitive. Most basic selection approaches mainly exploit signal features at single link level (e.g. carrier-to-noise power density ratio $C/N_0$, elevation angle $\theta$, etc.) so as to exclude -- or mitigate the influence of -- satellites %that have 
that would presumably contribute to large positioning errors~\cite{CN0SigmaE,CN0SigmaDelta}. 

For the remaining satellites fulfilling these basic single-link quality criteria, many selection techniques have been proposed such as, subset-testing \cite{subsetTesting}, RANSAC \cite{RANSAC}, iterative reweighting \cite{iterativeReWeighting}, etc. These methods involve different tradeoffs between computational complexity and performance. However, because of the huge combinatorial complexity of testing all possible subsets of satellites, exhaustive search is not tractable and \textcolor{black}{this selection problem remains an open issue to the best of our knowledge. For instance, \textcolor{black}{most conventional} selection approaches (e.g., \cite{RAIMcon}) rely on the spatial distribution of intermediary positioning results conditioned upon specific subsets of the available satellites to determine the most harmful contributions through posterior consensus. However, they mainly exclude satellites with strongly biased pseudo-ranges, which may still be insufficient %for achieving accurate positioning without mitigating the inaccuracy affecting selected pseudo-ranges. 
to draw the best possible accuracy out of selected pseudo-ranges, given that their respective -- and even joint -- negative influence is not properly mitigated.
}
%\subsection{our contribution}
In this paper, we introduce a novel pre-processing approach suited to single-epoch stand-alone positioning, which aims at overcoming major drawbacks of conventional selection techniques.
%\textcolor{black}{\textit{[V: isn't this discussion redundant to the above paragraph? Let's say above all about previous methods and focus here only on our method and not go back.]}The latter indeed mainly rely on link-wise satellite features or metrics (by means of calibrated empirical functions) on the one hand, and suffer from high online computational complexity (due to the inherent combinatorial nature of the hard selection problem) on the other hand.}
More specifically, %is intended to overcome these limitations by 
we redefine the initial satellites selection problem as a weighting problem, where one first originality lies in the application of \ac{ML} into the domain/space of pseudo-range residuals (i.e., relying on intermediary positioning results conditioned on specific subsets of satellites) for determining the best satellite weights. 

In order to fully exploit the potential of deep learning tools in harnessing hidden correlations between pseudo-range measurements, as well as possible joint effects from discarding several satellites at a time (on final positioning performance), we exploit a \acf{LSTM NN}, which is fed with a customized pseudo-range residuals matrix (processed as a whole), representing a second originality of our contribution. This NN is trained to predict quality factors that account for the link-wise standard deviations of pseudo-range errors. These predictions are finally used to compute nearly-optimal satellite weights within a standard \ac{WLS} positioning solver. %\textcolor{black}{The main paper novelty hence lies in i)..., ii)..., and iii).} 

To sum up, our main contributions are two fold. First, we introduce a novel deep learning-based technique to solve the satellite selection problem dedicated to improve the accuracy of single-epoch positioning. The main ingredients of our approach are the \ac{LSTM NN} architecture coupled with the new and customized pseudo-range residual matrix used as the NN input. Second, we improve the computation of the measurement weights in comparison with conventional parametric methods.\\
Finally, our approach is tested on both synthetic simulation data and real-field experimental data from extensive measurement campaigns, which were conducted with a dedicated test platform under typical vehicular mobility in a variety of scenarios and environments.

Note that our approach can be beneficial to both real-time location-based applications requiring accurate positioning information (e.g., autonomous vehicles, unmanned aerial vehicles...) and offline post-processing applications (e.g., aiming at correcting raw online \ac{GNSS} trajectory a posteriori).

%In the following sections, 
The rest of this paper is structured as follows. First, Section~\ref{sec:problem_formulation} introduces the system model and the general problem formulation. On this occasion, we also recall representative satellites selection techniques from the literature. Then, our construction of the residuals matrix, its use as an input to the \ac{LSTM NN}, as well as the underlying machine learning model, are detailed in Section~\ref{sec:proposed_system_architecture}. Finally, numerical results on both synthetic and real datasets are analyzed in Section~\ref{sec:results}.

\section{Problem Formulation}\label{sec:problem_formulation}

We start by describing the problem under study and then we summarize the most representative existing work.

\subsection{Single epoch solution}

For the sake of simplicity and without any loss of generality, we consider a set of $N$ single band, single constellation, pseudo-ranges  measurements $\{\rho^{i}\}_{i=1\hdots N}$ (the same formulation would apply for carrier phase, pseudo-range rates, etc.), while for the experimental validations reported in Sec.~\ref{sec:results} we will deal with multi-band and multi-constellation scenario. For those pseudo-range measurements, the appropriate compensations computed from ephemeris data (\ac{SV} clock bias, ionospheric and tropospheric delays, Sagnac correction etc.)  have already been applied. In the absence of multi-path or any strong bias, the $i$-th \ac{SV} measurement can be modeled as  
\begin{equation}
    \rho^{i} = \sqrt{(x-x^i)^2+(y-y^i)^2+(z-z^i)^2} + c\ \delta + \eta^i,
\end{equation}
where $\rho^{i}$ is the pseudo-range between a receiver $R$ and the \\$i$-th $\ac{SV}$,  and  $(x^{i},y^{i},z^{i})$ and $(x,y,z)$ the coordinates for the \\$i$-th satellite and the receiver, respectively. The parameter $c$ is the speed of light, and $\delta$ is the clock bias between the receiver and the considered constellation; $\eta^i$ is the observation noise which represents the receiver noise and residual errors from ionosphere and troposphere delays, etc. Although ionosphere and troposphere residual errors (i.e. after correction from navigation message)  are highly correlated over time, they could be considered as independent and zero mean for single epoch processing. Hence, we can  assume that the observation noise follows a centered Gaussian distribution  $\eta^i \sim \mathcal{N}(0,\,\sigma_{i}^{2})$.

Our aim is to estimate the vector $\ve{X}=[x,y,z,\delta]^\top$ from the measurements $\{\rho^{i}\}_{i=1 \hdots N}$. A widely used and efficient solution is provided by the maximum-likelihood estimator (MLE) \cite{Rossi2018}, which simplifies to a weighted least-squares for our Gaussian noise model
\begin{equation}
 \label{eq:MLE_Result}
 \hat{\ve{X}}= \arg \min_{\ve{X}} \ \sum_{i=1}^{N}\omega^{i}(\rho^{i}-h^{i}(\ve{X}) \big)^2,
 \end{equation}
with the observation function for satellite $S_i$ defined as 
\begin{equation}
 \label{eq:Observation_function}
 h^{i}(\ve{X}) = \sqrt{(x-x^i)^2+(y-y^i)^2+(z-z^i)^2} + c \ \delta,
 \end{equation}
and the weights equal to 
\begin{equation}
 \label{eq:WeightsDefinition}
 \omega^{i}=\frac{1}{(\sigma^2)^i}.
 \end{equation}
The solution can be easily computed using an optimization algorithm such as Gauss-Newton or Levenberg-Marquardt \cite{Levenberg-Marquardt_Algo}.

\subsection{\ac{GNSS} Satellite Selection and Weighting Problems}

%\ac{GNSS} satellite selection is a complex process that usually involves a combinatorial search over all the available satellites based on their respective signal characteristics, such as $C/N_0$ and elevation angle. The problem is further complicated by the dynamic nature of the environment, such as the presence of obstacles, interference, and signal fading, which can affect the quality of the received signals. Beyond, the availability of multiple \ac{GNSS} constellations and the increasing number of available satellites exacerbate the problem by increasing \textcolor{black}{exponetially} the number of possible satellite combinations. 

Generally, a first basic \ac{SV} selection based on satellite elevation or $C/N_0$ thresholds is performed to excluded presumably strongly biased measurements.
Then, the standard deviation of the remaining measurements will be estimated using an empirical functions, for example, as the following,
\begin{equation}
\label{eq:SoTA_NoiseDef}
(\sigma^2)^{i}=\frac{1}{\sin^2{(\theta^i)}}\left(\sigma^2_{\rho Z}+\frac{\sigma^2_{\rho c}}{(C/N_0)^i}+\sigma^2_{\rho a} (a^2)^i\right),
\end{equation}
where this functions mainly depends on satellite elevation $\theta_0$, $C/N_0$, acceleration $a^i$, and other empirically calibrated coefficients ($\sigma^2_{\rho Z}, \sigma^2_{\rho c}, \sigma^2_{\rho a}$) that are hard to be tuned.

 However, some measurements could be strongly biased by multi-path for instance, and violate the expected Gaussian model resulting in a significant degradation of the solution accuracy. It is thus of primary importance to exclude these measurements from the solution, either by discarding them or by assigning them a zero weight, which is called de-weighting. Such measurements can be efficiently detected at the navigation processor stage based on innovation monitoring tests for instance \cite{EKFInnovation}, but this requires that the tracking filter has converged and that the predicted state (i.e., position, receiver clock offset, etc.) is accurate enough.  
Nevertheless, for single epoch processing, no predicted solution is available and \ac{SV} selection relies on measurements only with very limited prior knowledge. This implies that the detection of $k$ faults among $N$ measurements could potentially results in a huge number of subsets, $C_k^{N}$, to test in case of an exhaustive search, which may be intractable in real-time and even for post-processing. As an example, assuming at most 10 faults among 40 measurements would result in more than $847\times10^6$ subsets to test, which is computationally prohibitive.

\subsection{Existing Works}\label{subsec:Existing_work}

In \ac{GNSS}, \ac{FDE} usually rely on statistical tests and consistency checks to identify and reject corrupted signals to improve the integrity of the navigation solution. Various \ac{FDE} techniques already exist in the literature, such as classical \ac{FDE} \cite{knight2009comparison}, \cite{kuusniemi2007user}, the brute force subset testing approach \cite{angrisano2013gnss}, ARAIM techniques \cite{zhai2018fault}, \cite{joerger2014solution}, and others that depend on the Range Consensus (RANCO) \cite{RAIMcon}. More recent works \cite{VTCPaper2021} have proposed a new \ac{FDE} algorithm that is based on both a standalone \ac{FDE} block \textcolor{black}{making use of the residual test relying on \ac{WLS},} and an \ac{FDE}-based Extended Kalman Filter (EKF). This solution alternates between the two branches, based on a covariance matrix threshold. Typically, the EKF utilizing \ac{FDE} is employed when the covariance matrix falls below a pre-defined threshold. If the covariance exceeds this limit, the \ac{FDE} is used on its own. This algorithm was shown to provide significant performance gains in terms of accuracy compared to conventional state-of-the-art \ac{FDE} algorithms. 
\textcolor{black}{For this reason, we choose it as a reference for benchmark purposes in this paper.} However, since we consider by definition a single-epoch localization application, we have configured this algorithm to use only its standalone \ac{FDE} block. 

\section{Proposed System Architecture}\label{sec:proposed_system_architecture}
As mentioned above, we herein consider a single-epoch stand-alone positioning framework based on pseudo-ranges, \textcolor{black}{ without performing differential corrections}, where measurements are just pre-processed based on the ephemeris data (i.e., compensating for ionospheric, tropospheric, and Sagnac effects, as well as satellite clock errors).
One main objective is hence to leverage the complex -- and likely hidden -- inter-dependencies and joint effects (over multiple links) through supervised deep machine learning, so as to efficiently weight (or de-weight) the contributions from all satellites. 

\begin{figure*}
\resizebox{1\textwidth}{!}{\includegraphics{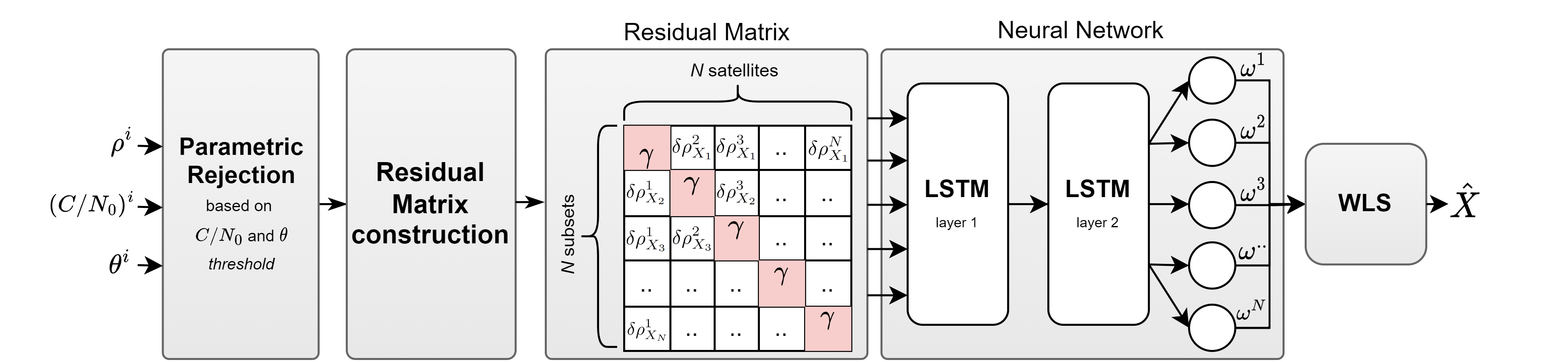}}
\caption{The complete architecture of our proposed approach.}
\label{fig_ApproachDiagram}
\end{figure*}

Overcoming the difficulty of labeling the data per link on the one hand, while still observing the joint effects on positioning performance of multiple links from various standpoints (e.g., link quality, Geometric Dilution of Precision (GDoP), etc.) on the other hand, \ac{ML} is thus applied directly into the space/domain of pseudo-range residuals. For this sake, our approach consists in constructing a matrix of such residuals, \textcolor{black}{where the $i$-th row contains all the residuals computed from a solution while excluding the $i$-th measurement.} Then, using this matrix as the input of a \ac{LSTM NN}. The latter is trained to predict \textcolor{black}{ the weights $\hat{\omega}^i$} related to the underlying distribution of pseudo-range errors according to (\ref{eq:WeightsDefinition}) for valid measurements, noting that a single measurement is observed from each distribution $ \mathcal{N}(0,\,\sigma_{i}^{2})$. In addition, we expect that the algorithm will predict nearly null weights $\hat{\omega}^i \approx 0 $ for strongly biased satellites to exclude them. 
%in the final \ac{WLS} positioning stage \textcolor{blue}{as the following. 
%\begin{equation} \label{OptWeights}
 %   \omega^{S_i} = \hat{\sigma}^{S_i}.
%\end{equation} 
Accordingly, we turn the initial (hard) selection problem into a (soft) weighting problem. The overall system architecture of our proposed approach is illustrated in Fig.~\ref{fig_ApproachDiagram}. 

\subsection{Residual Matrices Construction}\label{subsec:Residual_matrices_construction}
At each navigation epoch, we assume that multiple ($N$) satellite signals are received and a new matrix of positioning residuals \textcolor{black}{$\mathbf{M}$} is constructed, as follows. We generate $N$ subsets \textcolor{black}{$S_n$} of $N-1$ satellites, where we exclude one distinct satellite \textcolor{black}{(i.e, $n^{th}$ satellite)} at a time.
\begin{equation}
\label{eq:SubsetDef1}
S_n=\{\rho^i\},~i=1 \hdots N,~i\neq n
\end{equation}

For each subset $S_n$, we calculate the corresponding solution $X_{n}$  using (\ref{eq:MLE_Result}). Then, for each of the $N$ resulting positions $\{\ve{X}_{1},..., \ve{X}_{N}\}$ we calculate the $N-1$ pseudo-range residuals 
\begin{equation}
\label{eq:SubsetDef2}
\delta\rho_{{X}_n}^i=\rho^i-h^i(\ve{X}_n),~ ~i\neq n
\end{equation}
Coefficient $[\mathbf{M}]_{n,i}$ (i.e. row $n$, column $i$) of the residual matrix $\mathbf{M}$ is simply given by the corresponding residual for non-diagonal coefficient, or by an arbitrary large value $\gamma$ for the diagonal terms indicating that the satellite has been excluded (the pseudo-code for constructing the residual matrix is shown in Algorithm \ref{alg_ResMat})
\begin{equation}
\label{eq:SubsetDef3}
[\mathbf{M}]_{n,i}=
\begin{cases}
      \delta\rho_{X_n}^i,~ ~i\neq n \\
      \gamma, ~i=n
\end{cases} 
\end{equation}
Each row $n$ of the matrix will thus provides residuals associated with the exclusion of the $n
$-th measurement. \textcolor{black}{Although it assumes a single fault per subset (i.e., row), the motivation of building such a matrix is that it may be able to reveal hidden inter dependencies between the measurements and their effects on the computed solution, while being fed as a single input to the neural network}.  
%\textcolor{blue}{$\{r_{ni}, i\in[1..n[\cup ]n..N]\}$} associated with the non-excluded satellites, while artificially associating the entry corresponding to the excluded satellite \textcolor{blue}{$\{r_{ni}, n=i\}$} with a high value, accounting for an arbitrary large pseudo-range residual \textcolor{blue}{to allow the \ac{ML} model to distinguish between the missing and present residuals}. Therefore, the generated matrix is of dimension $N\times N$. 

%A pseudo-range residual ($r_{ni}$) is calculated as: 
%\begin{equation}
 %   \textcolor{blue}{r_{ni}=|\rho_R^{S_i}-\xi^{S_i}_{\hat{R}_n}|,}
%\end{equation}
%\textcolor{blue}{where the $\rho_R^{S_i}$ is the pseudo-range measured by the receiver, and $\xi^{S_i}_{\hat{R}_n}$ is the the distance between the $n^{th}$ estimated receiver position and the satellite position. The output labels, which are later used in the supervised learning phase, are generated as the absolute values of the differences between measured pseudo-ranges and actual ground-truth ranges.
%where as the absolute difference between the pseudo-range measured by the receiver and the distance between the estimated receiver position and the satellite position. The output labels, which are later used in the supervised learning phase, are generated as the absolute values of the differences between measured pseudo-ranges and actual ground-truth ranges.

\SetKwInput{KwData}{Input}
\SetKwInput{KwResult}{Output}
\RestyleAlgo{ruled}
\DontPrintSemicolon
\begin{algorithm}
\caption{Residual Matrix construction}\label{alg_ResMat}
%\KwData{$\rho^{S_i}_R, X^{S_i} (i=1..N)$}
%\KwResult{Residual Matrix ($M$) }

\For{$n \in \mathrm{range}(N)$}{
$\ve{X}_n \leftarrow \arg\min_{{\substack{\\\ve X}}}\big(\sum_{\substack{i=1 \\ i \neq n}}^{N}(\rho^i-h^{i}(\ve{X}) \big)^2\big)$ \;

\For{$i \in \mathrm{range}(N)$}{
\eIf{n=i}{$\mathbf{M}$[n,i] $\leftarrow \gamma$}
{$\mathbf{M}$[n,i] $\leftarrow \delta\rho_{X_n}^i$ }
}
}
\Return{$\mathbf{M}$}
\end{algorithm}

\subsection{Long-Short Term Memory Neural Network}

The overall structure of the proposed \textcolor{black}{residual} matrix and more particularly, the evolution of the pseudo-range residual patterns over the $N$ subsets (i.e., over the matrix rows), both provide rich information about the sought inter-dependencies and their combined effects, which can be advantageously captured by machine learning tools. By analogy, the pseudo-range residuals matrix can be interpreted as a concatenation of $N$ feature column vectors, each containing the feature values for $N$ pseudo time steps. Each feature column vector thus corresponds to the set of pseudo-range residuals of one single satellite from each of the $N$ subsets. 

In this kind of problems, the \ac{LSTM NN} algorithm, which is a type of \ac{RNN} \cite{huang2019deep}, has the advantage of keeping memory over multiple (possibly distant) pseudo time steps. Hence, is also suited to exploit the correlations across the matrix rows in our case, even if we explicitly deal with a single-epoch problem. Similar applications of the \ac{LSTM NN} to other time-invariant problems have already been considered. For instance, in \cite{HandwritingRec}, \ac{LSTM NN} was used to process data with long-range interdependence (i.e., using geometric properties of the trajectory for unconstrained handwriting recognition). 

\section{Numerical Results}\label{sec:results}

\textcolor{black}{Here, we analyse the performance of our proposed deep learning based approach obtained on synthetic and real data.}

\subsection{Synthetic Data}
In order to thoroughly test and validate the feasibility of our proposed approach in capturing the hidden inter-dependencies within the residual matrix, but also its ability to accurately approximate the standard deviations of pseudo-range errors,  %The predicted standad deviation is used in the weights of the Weighted Least Squares (WLS) algorithm, 
we first generate a dataset of synthetic simulation data mimicking the behavior of a real \ac{GNSS} systems. This allows us to illustrate the main performance trends, as well as to validate our approach before applying it to real-world data (See Section~\ref{subsec:Real_data}).

This dataset consists of $180,000$ epochs, each containing $N=60$ pseudo-range measurements with respect to $60$ different satellites. The random noise terms affecting those pseudo-ranges are supposed to be independent and identically distributed (i.i.d.), following a distribution obtained as a mixture of normal and exponential distributions. The latter shall indeed account for the possibility to experience either typical errors expected from so-called ``good'' satellites or much more harmful (and likely positively biased) outlier measurements from ``bad'' satellites %\textcolor{black}{[Sorry but we're talking about the noise pdf here, so the symbol chosen for the density function cannot be the same as that used previously for the observation noise realization itself ($\eta$)!] }:
\begin{equation}
    f(x) = \alpha \frac{1}{\sigma \sqrt{2 \pi}} e^{ -\frac{(x - \mu)^2}{2\sigma^2} } + (1-\alpha) \lambda e^{-\lambda x},
\end{equation}
where $\alpha$ is the mixture parameter, which represents the probability that the measurement error realization is drawn from the normal distribution of mean $\mu$ and standard deviation $\sigma$, and $\lambda$ is the decay rate of the exponential distribution. All these parameters were fine-tuned so as to fit the empirical distribution of real pseudo-range measurements, which were collected along with a reference ground-truth system in various operating conditions. The latter will be used in Subsection \ref{subsec:Real_data} for further performance assessment. 

The generated synthetic data was then divided into three disjoint datasets: training, validation, and testing, with respective proportions of $60\%$, $20\%$, and $20\%$ \cite{handsOnML}, in order to ensure a robust evaluation of the \textcolor{black}{performance of our approach}. 
%our model performance. 
The process of dividing the data into three sets is a common practice in machine learning, known as the train-validation-test split. The training dataset is used to train the neural network and learn the underlying patterns in the data. The validation dataset is used to evaluate the performance of the neural network during training, and fine-tune the hyperparameters based on its performance. The testing dataset is used as an independent measure to evaluate the generalization performance of the prediction, which is an indication of how well the neural network will perform on new unseen data. This technique helps to prevent from overfitting effects, which occur when a neural network fits too closely the training dataset but performs poorly on new unseen data.

\begin{figure}
\resizebox{0.5\textwidth}{!}{\includegraphics{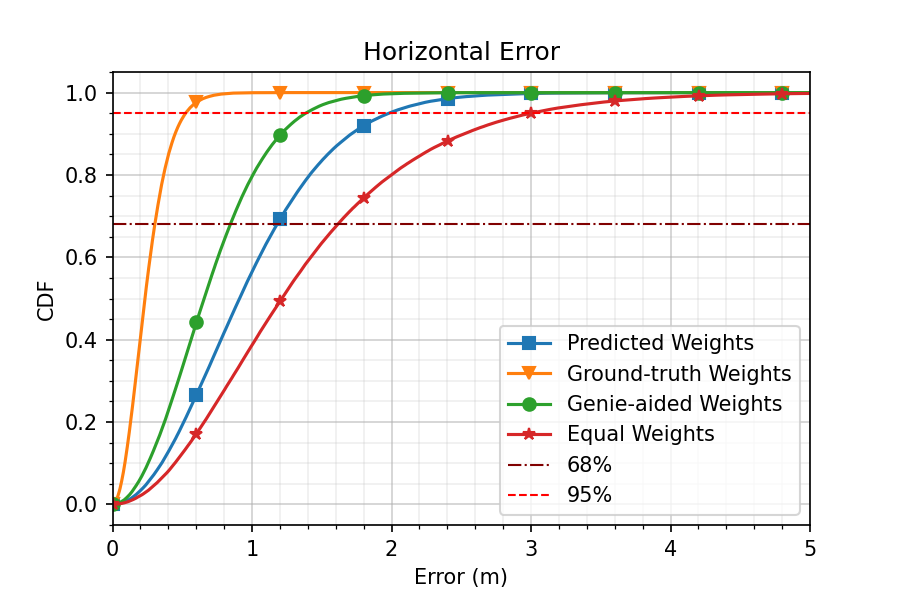}}
\caption{Empirical \ac{CDF} of horizontal positioning error for various measurements weighting methods, based on synthetic data.}
\label{fig_ArtificialData_9HorError}
\end{figure}

\begin{figure}
\resizebox{0.5\textwidth}{!}{\includegraphics{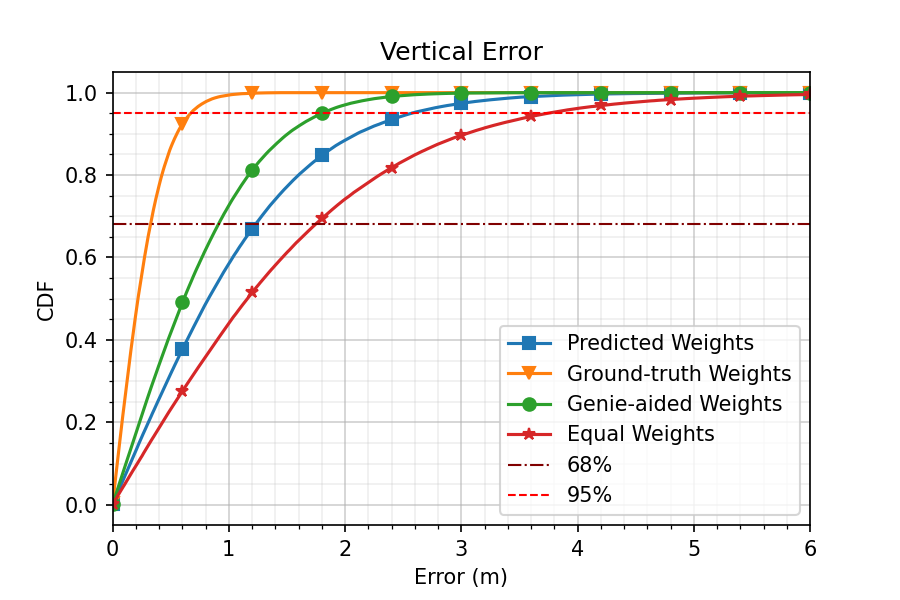}}
\caption{Empirical \ac{CDF} of vertical positioning error for various measurements weighting methods, based on synthetic data.}
\label{fig_ArtificialData_9VerError}
\end{figure}

\textcolor{black}{Note that labelling in our decision problem is a very challenging task since finding the best satellites' set is not tractable due to prohibitive combinatorial complexity (even offline and knowing the reference). This is the reason why we decided to convert the initial decision problem into a weighting problem. Following a supervised training, the data was hence labeled as:
\begin{equation}\label{eq:Weights_truth}
    \textcolor{black}{\omega^{i}=1/(\rho^{i}-h(\ve{X}_{true}))^2},
\end{equation}
where $\ve{X}_{true}$ is the ground truth position.
This intuitive choice is validated by our performance curves in Fig.~\ref{fig_ArtificialData_9HorError} and ~\ref{fig_ArtificialData_9VerError}.}

Using this synthetic dataset, an empirical evaluation was first conducted on \textcolor{black}{multiple neural network architectures}, including convolutional neural networks (CNN), fully connected neural networks (FCNN), and various other types of recurrent neural networks (Simple-\ac{RNN}, LSTM, Bi-LSTM and Gated recurrent units (GRU)). The simulation results then confirmed that the LSTM algorithm could provide the best \ac{WLS} positioning performance (i.e., after applying the best weights out of NN predictions) in our problem. Its architecture was hence further optimized \textcolor{black}{empirically} during the training phase, based on real data. The best neural network consisted of one hidden LSTM layer containing $512$ neurons followed by a dense layer containing $60$ neurons with a ReLu activation function. On this occasion, the \ac{MSE} was used as loss function, and an early stopping callback was implemented to prevent overfitting. \textcolor{black}{The used neural weights optimizer} during training was \emph{Adam} \textcolor{black}{\cite{adam}}. 

Finally, the \ac{WLS} positioning performance based on our prediction method has been compared with theoretical bounds (i.e., \ac{CRLB}) \textcolor{black}{\cite{CRLB}}, as well as two genie-aided solutions %endowed with so-called optimal weights (i.e., with perfect prior knowledge of the ground-truth position or of the good satellites, resp.)
on the one hand, and a suboptimal unweighted strategy on the other hand (i.e., as a worst-case baseline solution with equally-weighted pseudo-ranges from all  the satellites). More precisely, the first tested method, referred to as ``\textcolor{black}{Ground-truth Weights}'', 
%\footnote{This case is somehow unrealistic and definitely ideal. Its so-called ``optimality'' lies in the fact that, given one single observation per link per epoch, the absolute value of the actual pseudo-range error could be statistically interpreted as the best possible approximation to the standard deviation characterizing the actual underlying error density for this link.}
utilizes weights as in (\ref{eq:Weights_truth}) with the ground-truth position, before applying \ac{WLS} positioning. 
The second method, named ``Predicted weights'', uses \textcolor{black}{the proposed algorithm to predict link-wise quality factors which is used to compute weights as in  (\ref{eq:WeightsDefinition}) for \ac{WLS} positioning.}  %predicted through our proposed approach, which approximate the standard deviations of pseudo-range errors that are used to compute optimal weights in \ac{WLS} positioning.
The third method, referred to as the ``Genie-aided'' solution, is based on the assumption that a prior knowledge of the biased satellites is available, and utilizes this information to completely exclude biased satellites from the localization solution. This method serves here as a first reference in our benchmark, accounting for the best performance that could be achieved with perfect detection and exclusion of biased satellites.
Finally, in the fourth ``Equal weights'' method, all the satellite measurements are taken into account in \ac{WLS} positioning (i.e., with the same relative importance, regardless of their actual errors). This somehow represents a worst-case assumption \textcolor{black}{without any exclusion}, and serves as another reference in our benchmark, accounting for the performance that can be achieved without any prior knowledge, neither of the biases themselves, nor of their respective statistics.

Fig. ~\ref{fig_ArtificialData_9HorError} and \ref{fig_ArtificialData_9VerError} show the empirical cumulative density function (\ac{CDF}) of horizontal and vertical positioning errors obtained with the four weighting strategies above. Those methods were first applied on a dataset including $9\%$ of strongly biased satellites (i.e., among the $60$ available).
%
%Highlighting the improvement in accuracy achieved by our approach 
When compared to the evenly weighted positioning solution, the proposed approach is then shown to yield a typical accuracy improvement in $95\%$ of the tested epochs (i.e., \ac{CDF} at the characteristic $95\%$-quantile) of about $1.02$ m in terms of horizontal errors (See Fig.~ \ref{fig_ArtificialData_9HorError}) and $1.18$ m in terms of vertical errors (See Fig.~\ref{fig_ArtificialData_9VerError}). \textcolor{black}{On the other hand, when compared with the ''Genie-aided'' approach, the difference in accuracy for $95\%$ of the tested epochs is about $0.59$ m in terms of horizontal error, and $0.77$ m in terms of vertical error.}

\begin{figure}
\resizebox{0.5\textwidth}{!}{\includegraphics{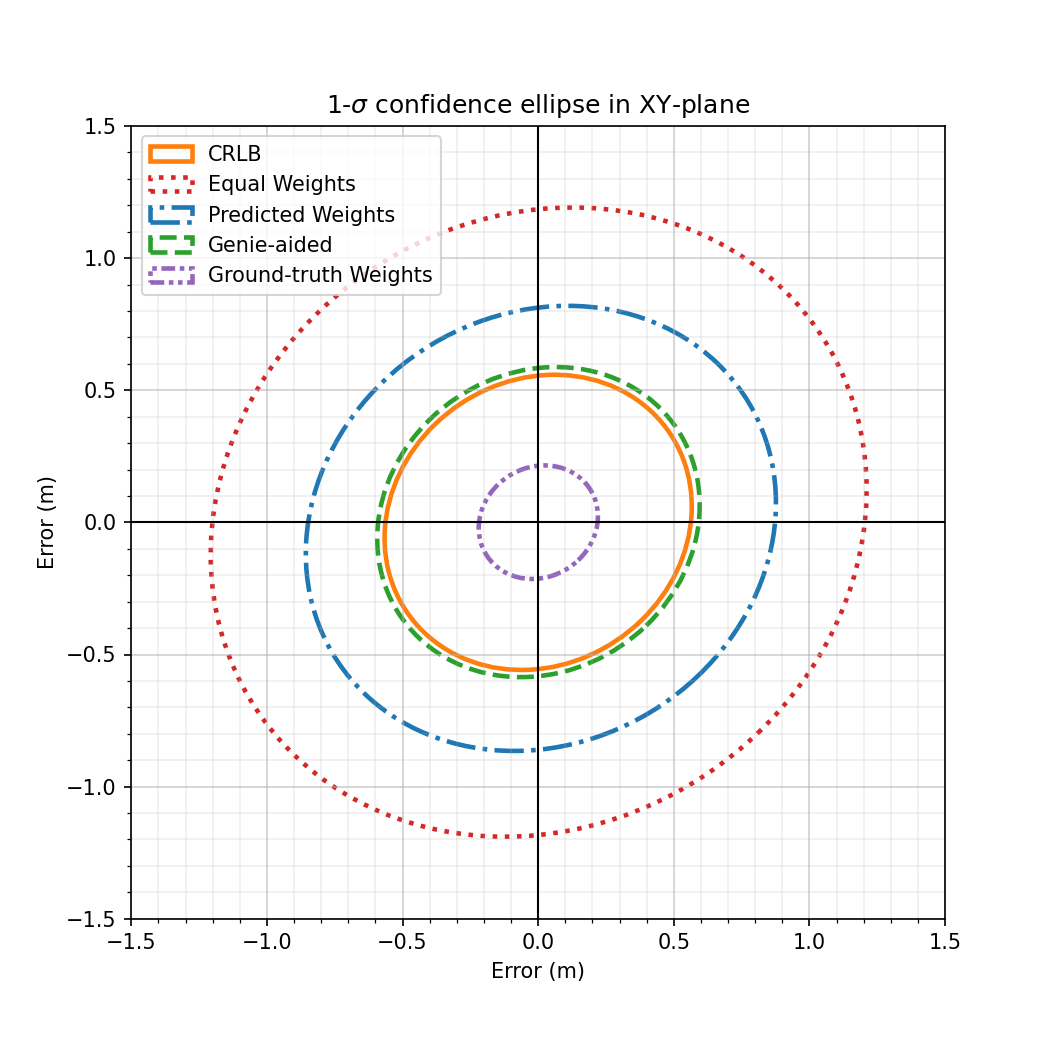}}
\caption{Comparison of Positioning confidence ellipses for various approaches with the \ac{CRLB}.}
\label{fig_ArtificialData_CRLB}
\end{figure}

\textcolor{black}{Fig. \ref{fig_ArtificialData_CRLB} presents the \textcolor{black}{1-$\sigma$} confidence ellipse for various weighting approaches in comparison with the \ac{CRLB}. The results are acquired through the execution of Monte Carlo trials for a specified satellite and receiver geometry. The "Genie-aided" weighting approach shows close agreement with the \ac{CRLB} in terms of accuracy, while the evenly weighted approach demonstrates inferior accuracy. The "Predicted weights" extracted from our approach, exhibits accuracy that lies between the \ac{CRLB} and the "Equal weights" with a tendency towards the \ac{CRLB}. The "Ground-truth Weights" approach outperforms the \ac{CRLB} since noise mitigation is based on a perfect knowledge of the exact bias, which would not be available to any real estimator.}

Still based on synthetic data, a sensitivity study was also conducted to analyze the performance of our approach for different percentages of strongly biased satellites. A preliminary analysis of the real dataset revealed that concrete real-life cases could have up to $9\%$ biased satellites (i.e., among all the available satellites per epoch). Accordingly, we generated different synthetic datasets, while varying the percentage of \textcolor{black}{strongly-biased} satellites (among the $60$). %from $1\%$ to $10\%$, as a worst-case scenario.

\begin{figure}
\resizebox{0.5\textwidth}{!}{\includegraphics{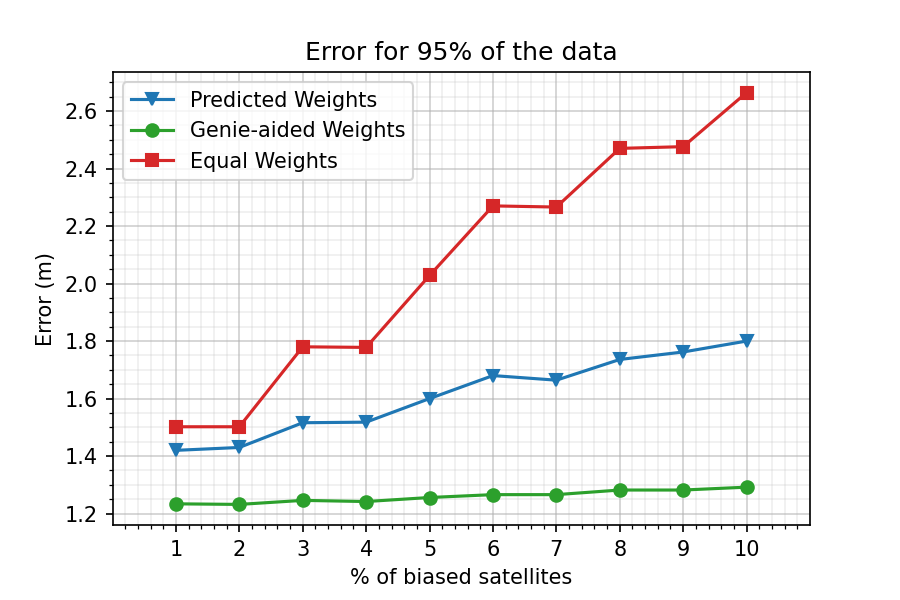}}
\caption{$95\%$-quantile of  positioning error for different measurements weighting methods, as a function of the ratio of biased satellites (among $60$), based on synthetic data.}
\label{fig_ArtificialData_AllPerc}
\end{figure}

%Our proposed approach was evaluated on a variety of datasets that varied in the proportion of biased satellites. The performance of our approach was compared to both an equally weighted approach and a genie-aided approach. 
Fig.~\ref{fig_ArtificialData_AllPerc} shows the evolution of the positioning error at $95\%$ of the \ac{CDF} (i.e., the $95\%$-percentile) as a function of this varying ratio.
The results indicate that, as the proportion of biased satellites increases, the performance gain achieved with our approach in comparison with the equally weighted approach tends to grow more rapidly than the performance degradation observed in comparison with the idealized ``Genie-aided weights'' approach, 
remaining in the same order of magnitude (i.e., less than $2$ m vs. more than $2.6$ m with the ``Equal weights'' strategy). 
This allowed us to further validate the robustness of our approach (i.e., far beyond the toy case of one single biased satellite), before applying it to real-world data, as discussed in the next section.

\subsection{Real Data}\label{subsec:Real_data}

%After the first validations above obtained based on synthetic data, 
\textcolor{black}{Here, we present the most representative results obtained by extensive experimental testing} and validation based on real-life data, which was collected from multiple measurement campaigns in a variety of operating conditions. These conditions included open skies, dense urban areas, and various mobility regimes. The data was collected using cutting-edge pieces of equipment and more specifically, an ublox ZED-F9P receiver, which is dual-band and endowed with Real-Time Kinematics (RTK) capabilities. A side cm-level ground-truth referencing system was also utilized to ensure the accuracy and integrity of the collected data, as well as to establish the ground-truth information. The tested \ac{GNSS} receiver is capable of receiving up to $N=60$ satellite signals from multiple \ac{GNSS} constellations (i.e. GPS, GLONASS, GALILEO, etc.) over multiple frequencies (i.e. L1, L2, E1, and E5 bands). Overall, this  testing phase consisted of $56$ distinct sessions for a total of $181,000$ epochs, providing a comprehensive,  representative and diverse dataset for evaluation.

\begin{figure}
\resizebox{0.5\textwidth}{!}{\includegraphics{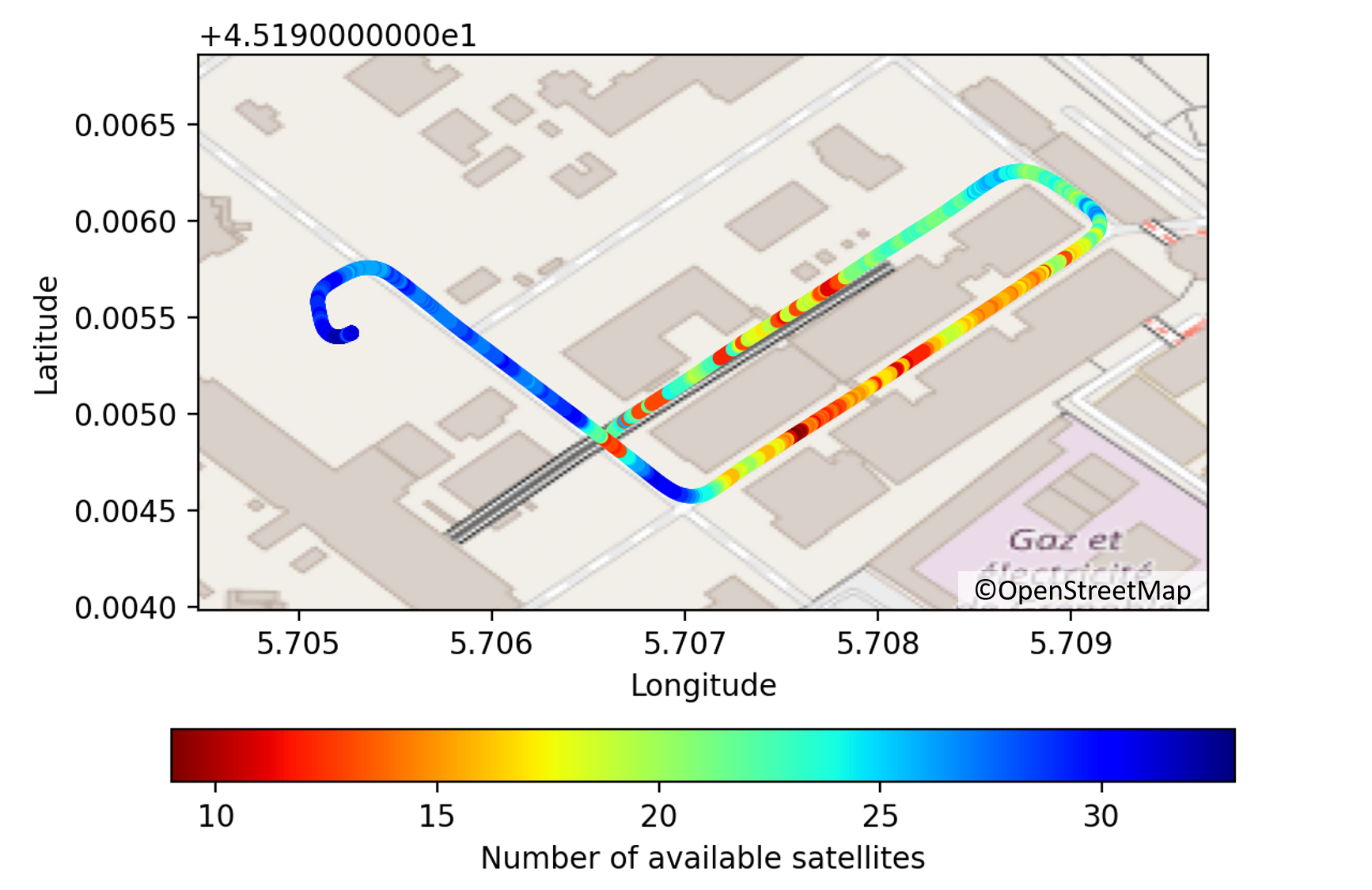}}
\caption{Variation of satellite availability while navigating in different conditions (open sky vs urban area).}
\label{fig_Map_SatelliteDensity}
\end{figure}

The number of satellites received during each epoch could fluctuate, depending on the operating conditions. In certain environments, such as crowded urban areas for instance, the number of available satellites could be significantly lower than in open sky conditions \textcolor{black}{as illustrated in Fig. \ref{fig_Map_SatelliteDensity}. The figure depicts the variation in satellite availability during navigation in two conditions: open sky and urban areas. During this session, the availability in open sky regions remains stable at approximately $30$ satellites, whereas in \textcolor{black}{narrow urban canyons such as the one depicted in Fig. \ref{fig_Map_SatelliteDensity}}, the availability drops to a minimum of $9$ satellites, as indicated by color-coded fluctuations. It is worth noting that in other sessions, the number of available satellites may increase up to $45$.} This variability is also influenced by the position of the satellites \textcolor{black}{ at the time of navigation}. As a result, the dimension of our residuals matrix could have significantly varied as well. To address this issue, we employed padding to resize all the residuals matrices to a common matrix size. Additionally, we leveraged the ability of LSTM networks to handle variable-length input sequences by masking the non-present steps of the sequence, represented by rows of the residuals matrix in our representation.

For the purpose of optimizing the neural network architecture with a minimal number of layers and neurons while preserving performance, we employed a technique known as grid search. This involved varying the number of hidden layers and the number of neurons per layer as hyperparameters. To prevent overfitting, we utilized an early stopping callback during the training process. The performance of each trained network was evaluated on an unseen test dataset. Through this process, we could determine that the optimal architecture consisted of $2$ hidden layers with $893$ neurons in each (See Fig.~\ref{fig_ApproachDiagram}). \textcolor{black}{Moreover, the neural network was trained using labels that were defined according to   (\ref{eq:Weights_truth}), where the $\ve{X}_{true}$ stands for the ground truth position collected from the reference ground-truth system.}

\begin{figure}
\resizebox{0.5\textwidth}{!}{\includegraphics{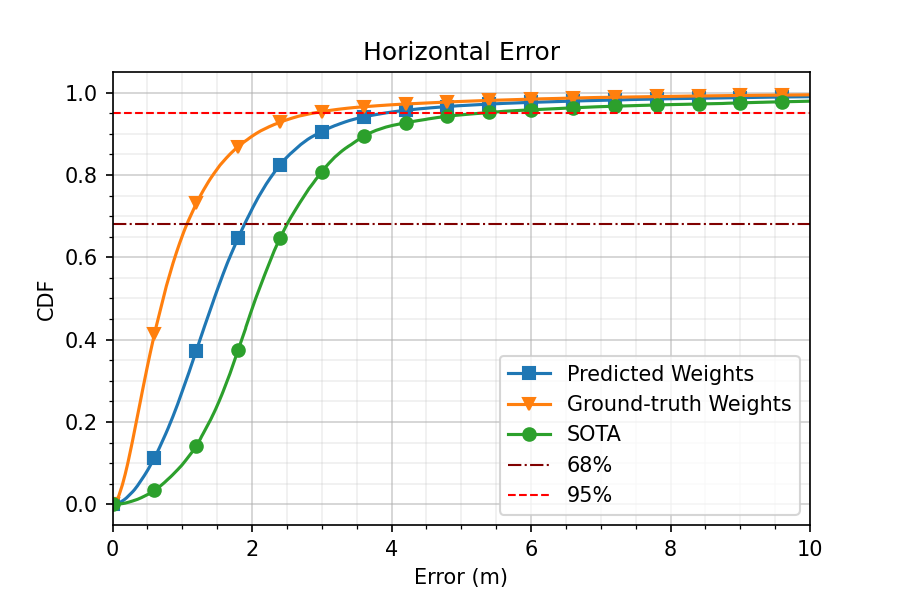}}
\caption{Empirical \ac{CDF} of horizontal error for various measurements weighting/selection strategies (incl. \cite{VTCPaper2021}), based on real field data.}
\label{fig_RealData_HorError}
\end{figure}

\begin{figure}
\resizebox{0.5\textwidth}{!}{\includegraphics{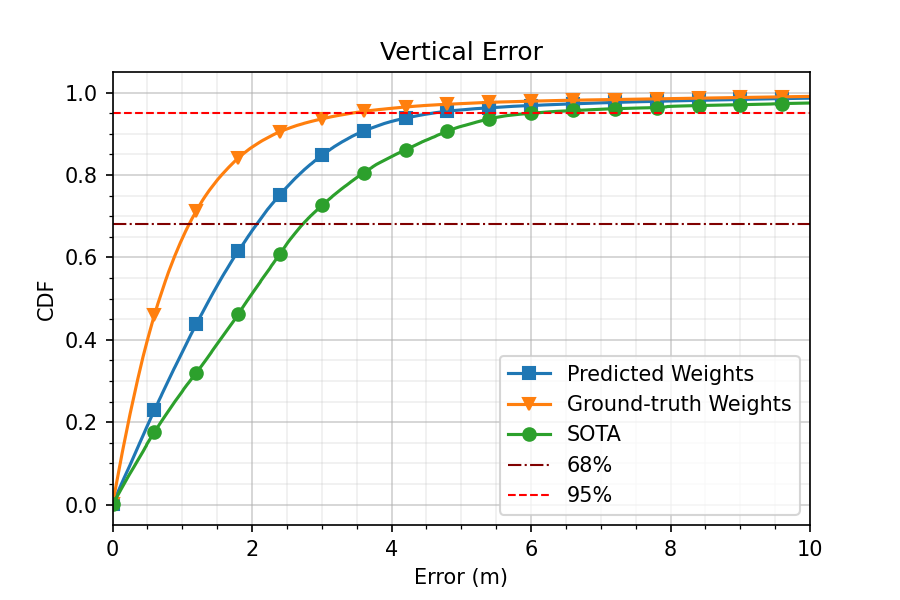}}
\caption{Empirical \ac{CDF} of horizontal error for various measurements weighting/selection strategies (incl. \cite{VTCPaper2021}), based on real field data.}
\label{fig_RealData_VerError}
\end{figure}

Similar to the tests performed on synthetic data, Fig.~ \ref{fig_RealData_HorError} and \ref{fig_RealData_VerError} 
show quite significant improvements in terms of both horizontal and vertical errors, in comparison with the state-of-the-art solution from \cite{VTCPaper2021} (See Subsec. ~\ref{subsec:Existing_work}). Typically, our approach %, which utilizes weighting of satellite measurements, 
exhibits a performance gain of $0.61$ m (resp. $1.38$ m) in terms of horizontal error at $68$\% (resp. $95$\%) of the \ac{CDF}. As for the vertical error, we also observe an improvement of $0.66$ m (resp. $1.43$ m) at $68$\% (resp. $95$\%) of the \ac{CDF}. This consistent performance improvement across all data points and for the different problem dimensions illustrates again the robustness of our proposal in improving localization accuracy under various operating conditions.

\begin{figure}
\resizebox{0.5\textwidth}{!}{\includegraphics{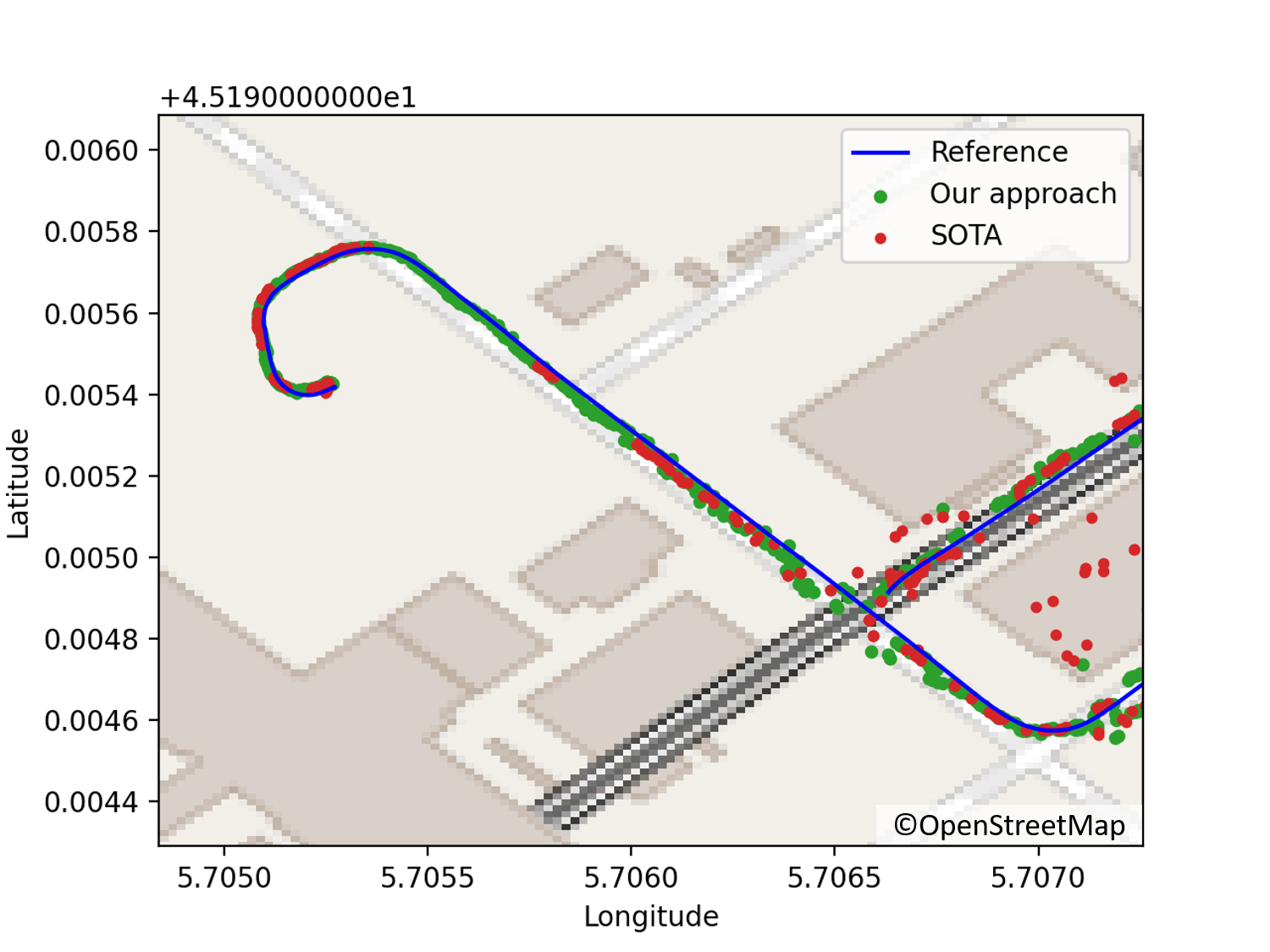}}
\caption{Example of positioning traces obtained for one of our field navigation sessions (urban canyon), for the proposed approach (green dots), the  approach in \cite{VTCPaper2021} from recent state-of-the-art (red dots), and the ground-truth reference system (blue solid line).}
\label{fig_MapCont}
\end{figure}

As \textcolor{black}{shown} in Fig.~\ref{fig_MapCont}, which depicts a segment of one navigation session in a particularly penalizing environment (urban canyon), our approach could provide  reliable single-epoch positioning results all along the tested trajectory,  whereas the state-of-the-art approach fails in several important portions. More generally, over all the tested sessions, the state-of-the-art approach was shown to fail in  providing any solution in about $8$\% of the tested epochs (in average), while our approach could systematically provide a reliable positioning solution. 
Our proposal thus seems even more particularly suited to severe operating conditions and/or challenging environments.

%The ability to consistently provide precise location information in challenging environments enhances the robustness and reliability of our approach. 

\section{Conclusions}

In this paper, we have introduced a new pre-processing technique for single-epoch standalone \ac{GNSS} positioning based on deep machine learning, which aims at optimally weighting the pseudo-range contributions from available satellites. In particular, we rely on an LSTM neural network architecture, which is fed by a customized matrix of conditional pseudo-range residuals.
Performance assessment on both synthetic data and real data resulting from multiple navigation sessions %suggest that our approach 
%could be all the more relevant and beneficial 
\textcolor{black}{show the high potential and relevance of our approach} in challenging operating contexts, where conventional parametric satellites selection techniques would fail. Accordingly, this solution is suited to real-time applications requiring good continuity of the navigation service, as well to offline applications necessitating high-accuracy traces retrieval.

Future works could consider utilizing other satellite features/metrics (e.g., $C/N_0$, elevation) as complementary information channels (i.e., besides the matrix of residuals currently in use) while feeding the learning process.  

\bibliographystyle{IEEEtran}
\bibliography{main}

\vspace{12pt}

\end{document}